\documentclass[preprint,showpacs,showkeys,preprintnumbers,amsmath,amssymb]{revtex4}
\usepackage{amssymb}
\usepackage[dvips]{graphicx}
\DeclareMathOperator{\sech}{sech}

\begin{document}

\title{New interaction solutions from Lax pair related symmetry of the Generalized fifth order KdV equation }

\author{ Xi-zhong Liu, Jun Yu, Bo Ren }

\affiliation{Institute of Nonlinear Science, Shaoxing University, Shaoxing 312000, China}
\begin{abstract}
The nonlocal symmetry of the generalized fifth order KdV equation (FOKdV) is first obtained by using the related Lax pair and then localized in a new enlarged system by introducing some new variables. On this basis, new B\"{a}cklund transformation is obtained through Lie's first theorem. Furthermore, the general form of Lie point symmetry for the enlarged FOKdV system is found and new interaction solutions for the FOKdV equation are explored by using classical symmetry reduction method.
\end{abstract}



\maketitle
\section{Introduction}
In nonlinear science, many effective methods, including Hirota's bilinear method\cite{hirota,huxingbiao}, Darboux transformations (DT) and the B\"{a}cklund transformations (BT)\cite{guchaohao,rogers}, inverse scattering transformation (IST)\cite{gelfand,segur}, symmetry analysis\cite{lie,olver,bluman,jlwu,xrhu,liuxizhong,liuxizhong11} etc., have been developed to study integrable nonlinear systems. Usually, using these methods one can obtain multiple-soliton solutions and many kinds of nonlinear waves such as the conoidal periodic waves, Painlev\'{e} waves. However, the interaction solutions between solitons and nonlinear waves are hard to be obtained by these traditional methods. Recently, Lou and his colleagues proposed the localization procedure to use nonlocal symmetry for finding interaction solutions and got many interesting results\cite{xiaonan,huxiao}.

As we known, using Lie point symmetries of a differential system, one can transform given solutions to new ones via finite transformation and construct group invariant solutions by similarity reductions. Unfortunately, nonlocal symmetry can not be used directly to construct exact solutions for nonlinear systems. To concur that obstacle, one needs to introduce more dependent variables to localize it in a new closed system. In this paper, we apply the localization procedure to construct exact solutions for the famous generalized fifth order KdV equation (FOKdV) in the form
\begin{equation}\label{5kdv}
u_t-\frac{2c}{3}u_{xxxxx}+\frac{20c}{3}uu_{xxx}+\frac{40c}{3}u_xu_{xx}-20cu^2u_x=0,
\end{equation}
with c as an arbitrary constant.

The paper is organized as follows. The Lax pair related nonlocal symmetry of the FOKdV equation is obtained in section 2 and it is further localized in a new prolonged system by introducing multiple new variables in section 3. On this basis, the finite group transformation theorem is also found by using Lie's first principle. In section 4, the general form of Lie point symmetry group for the enlarged FOKdV system is obtained and the similarity reduction solutions are explored by using the classical symmetry approach. The last section is devoted to a short summary and discussion.

\section{Nonlocal symmetry of FOKdV equation}
It is well known that the FOKdV equation possesses the Lax pair
\begin{equation}\label{lax1xx}
\phi_{xx}-u\phi=0,
\end{equation}
\begin{equation}\label{lax1t}
\phi_t-2c(\frac{1}{3}u_{xxx}-2uu_x)\phi+4c(\frac{1}{3}u_{xx}-u^2)\phi_x=0,
\end{equation}
which means the consistent condition $\phi_{xt}=\phi_{tx}$ produces the FOKdV equation \eqref{5kdv}.

A symmetry $\sigma_u$ of the FOKdV equation is  defined as a solution of its linearized equation
\begin{equation}\label{5kdvlin}
\sigma_{u,t}-\frac{2c}{3}\sigma_{u,xxxxx}+\frac{20c}{3}\sigma_u u_{xxx}+\frac{20c}{3}u\sigma_{u,xxx}
+\frac{40a}{3}\sigma_{u,x}u_{xx}+\frac{40c}{3}u_x\sigma_{u,xx}-40cuu_x\sigma_u-20cu^2\sigma_{u,x}=0
\end{equation}
that means Eq. \eqref{5kdv} is form invariant under the transformation
\begin{equation}\nonumber
u\rightarrow u+\epsilon\sigma_u
\end{equation}
 
To seek the nonlocal symmetry of the FOKdV equation related with the Lax pair, we write $\sigma_u$ in the form
 \begin{equation}\label{linearizu}
\sigma_u= X(x, t,u,\phi,\phi_x)u_x+T(x, t,u,\phi,\phi_x)u_t-U(x, t,u,\phi,\phi_x).
 \end{equation}
Substituting Eq. \eqref{linearizu} into Eq. \eqref{5kdvlin} and eliminating $u_t, \phi_{xx}$ and $\phi_t$
by Eqs. \eqref{5kdv}, \eqref{lax1xx}, \eqref{lax1t}, we obtain the determining equations for the function $X, T$ and $U$. Calculated by  computer, the final result is
\begin{equation}\label{nonlocal5kdv}
\sigma_u=(\frac{d_1}{5}x+d_4)u_x+(d_1t+d_2)u_t+\frac{2d_1}{5}u-d_3\phi\phi_x,
\end{equation}
where $d_i (i=1\cdots 4)$ are arbitrary constants.

It can be seen from Eq. \eqref{nonlocal5kdv} that not only the Lax pair related symmetry but also the Lie point symmetry can be obtained by this method.

\section{localization of the Lax pair related symmetry and the B\"{a}cklund transformation }

The nonlocal Lax pair related symmetry of the FOKdV equation can now be easily obtained by setting $d_1=d_2=d_4=0$, the result is
\begin{equation}\label{laxsym}
\sigma_u=-d_3\phi\phi_x.
\end{equation}

Apparently, the nonlocal symmetry \eqref{laxsym} can not be used directly to construct explicit solutions for the FOKdV equation. To localize it, we need to introduce new dependent variable
\begin{equation}\label{phi1}
 \phi_1\equiv \phi_x.
\end{equation}
So the system is prolonged and the corresponding symmetry equations are
\begin{subequations}\label{linear}
\begin{equation}
\sigma_{\phi,x}-\sigma_{\phi_1}=0,
 \end{equation}
\begin{equation}
 \sigma_{\phi,xx}-\sigma_u\phi-u\sigma_{\phi}=0,
\end{equation}
\begin{equation}\nonumber
\sigma_{\phi,t}-\frac{2c}{3}\sigma_{\phi}u_{xxx}-\frac{2c}{3}\phi\sigma_{u,xxx}+4c\sigma_{\phi}uu_x
+4c\phi\sigma_uu_x+4c\phi u\sigma_{u,x}
\end{equation}
\begin{equation}+\frac{4c}{3}\sigma_{\phi,x}u_{xx}+\frac{4c}{3}\phi_x\sigma_{u,xx}-4c\sigma_{\phi,x}u^2
-8c\phi_xu\sigma_u=0
\end{equation}
\end{subequations}
It can be easily verified that the solutions of \eqref{linear} has the form
\begin{subequations}\label{pointsy1}
\begin{equation}
\sigma_{u} =2\phi\phi_1,
\end{equation}
\begin{equation}
\sigma_{\phi} =\frac{1}{2}\phi p,
\end{equation}
\begin{equation}
\sigma_{\phi_1} =\frac{1}{2}(\phi_1 p+\phi^3),
\end{equation}
\end{subequations}
where the new variable $p$ is introduced as
\begin{equation}\label{px}
p_x = \phi^2,
\end{equation}
with compatibility condition
\begin{equation}
p_t= \frac{4c}{3}(\phi^2u_{xx}-4\phi\phi_1 u_x+4\phi_1^2u-\phi^2u^2).
\end{equation}

Solving the symmetry equation of \eqref{px}
\begin{equation}\label{linpx}
\sigma_{p,x}-2\phi\sigma_{\phi}=0,
\end{equation}
with \eqref{pointsy1}, we get the symmetry for $p$
\begin{equation}\label{sigp}
\sigma_p = \frac{1}{2}p^2.
\end{equation}

The form of $\sigma_p$ in \eqref{sigp} indicates that $p$ is a solution of Schwarzian form of the FOKdV equation
\begin{equation}\label{schwarzform}
3c\{p;x\}^2+2c\{p;x\}_{xx}-3\frac{p_t}{p_x}=0,
\end{equation}
with the Schwarzian derivative $\{p;x\}=\frac{p_{xxx}}{p_x}-\frac{3}{2}\frac{p_{xx}^2}{p_x^2}$.

The results \eqref{pointsy1} and \eqref{sigp} indicate that the Lax pair related nonlocal symmetry \eqref{laxsym} is successfully localized in the new enlarged system \eqref{5kdv}, \eqref{lax1xx}, \eqref{lax1t}, \eqref{phi1} and \eqref{px} and the corresponding Lie point symmetry vector takes the form
\begin{equation}\label{pointV}
V=2\phi\phi_1\partial_{u}+\frac{1}{2}\phi p\partial_{\phi}+\frac{1}{2}(\phi_1 p+\phi^3)\partial_{\phi_1}+\frac{1}{2}p^2\partial_p.
\end{equation}

Now, let us study the finite transformation group corresponding to the Lie point symmetry \eqref{pointV}, which can be stated in the following theorem.

\noindent\emph{ \textbf{Theorem 1.}}
If $\{u,\phi,\phi_1, p\}$ is a solution of the prolonged system \eqref{5kdv}, \eqref{lax1xx}, \eqref{lax1t}, \eqref{phi1} and \eqref{px}, then so is $\{\hat{u}, \hat{\phi},\hat{\phi_1},\hat{p}\}$ with
\begin{subequations}\label{pointsy}
\begin{equation}
\hat{p} =-\frac{2p}{\epsilon p-2},
\end{equation}
\begin{equation}
\hat{\phi} = -\frac{2\phi}{\epsilon p-2},
\end{equation}
\begin{equation}
\hat{\phi_1} = \frac{2\epsilon\phi^3}{(\epsilon p-2)^2}-\frac{2\phi_1}{\epsilon p-2},
\end{equation}
\begin{equation}
\hat{u}= u+\frac{2\epsilon^2\phi^4}{(\epsilon p-2)^2}-\frac{4\epsilon\phi_1\phi}{\epsilon p-2},
\end{equation}
\end{subequations}
with arbitrary group parameter $\epsilon$.

\emph{Proof:} Using Lie's first theorem on vector \eqref{pointV} with the corresponding
 initial condition as follows
\begin{eqnarray}
\\ \frac{d\hat{p}(\epsilon)}{d\epsilon}&=& \frac{1}{2}\hat{p}(\epsilon)^2,\,\quad \hat{p}(0)=p,\\
\frac{d\hat{\phi}(\epsilon)}{d\epsilon}&=& \frac{1}{2}\hat{\phi}(\epsilon)\hat{p}(\epsilon),\,\quad \hat{\phi}(0)=\phi,\\
\frac{d\hat{\phi_1}(\epsilon)}{d\epsilon}&=&\frac{1}{2}(\hat{\phi_1}(\epsilon) \hat{p}(\epsilon)+\hat{\phi}(\epsilon)^3),\,\quad \hat{\phi_1}(0)=\phi_1,\\
\frac{d\hat{u}(\epsilon)}{d\epsilon}&=&2\hat{\phi_1}(\epsilon)\hat{\phi}(\epsilon),\,\quad \hat{u}(0)=u.
\end{eqnarray}
one can easily obtain the solutions of the above equations stated in Theorem 1, thus the theorem is
proved.
\section{new symmetry reductions of the FOKdV equation}
For the prolonged closed system \eqref{5kdv}, \eqref{lax1xx}, \eqref{lax1t}, \eqref{phi1} and \eqref{px}, we seek the general Lie point symmetry in the form
 \begin{equation}\label{vectorv1}
V=X\frac{\partial}{\partial x}+T\frac{\partial}{\partial
t}+U\frac{\partial}{\partial u}+\Phi\frac{\partial}{\partial \phi}+\Phi_1\frac{\partial}{\partial \phi_1}+P\frac{\partial}{\partial p},
\end{equation}
which means that the prolonged system  is invariant under the following transformation
\begin{equation}
\{x,t,u,\psi,\phi,\psi_1,\phi_1,p\} \rightarrow \{x+\epsilon X,t+\epsilon T,u+\epsilon U,\phi+\epsilon \Phi,\phi_1+\epsilon \Phi_1,p+\epsilon P\},
\end{equation}
with the infinitesimal parameter $\epsilon$.
Equivalently, the symmetry in the form \eqref{vectorv1} can be written as a function form
\begin{subequations}\label{sigmasy}
\begin{equation}
\sigma_{u} = Xu_{x}+Tu_{t}-U,
\end{equation}
\begin{equation}
\sigma_{\phi} = X\phi_{x}+T\phi_{t}-\Phi,
\end{equation}
\begin{equation}
\sigma_{\phi_1} = X\phi_{1x}+T\phi_{1t}-\Phi_1.
\end{equation}
\begin{equation}
\sigma_{p} = Xp_{x}+Tp_{t}-P.
\end{equation}
\end{subequations}
Substituting Eq. \eqref{sigmasy} into Eqs. \eqref{5kdvlin}, \eqref{linear}, \eqref{linpx} and eliminating $u_{t}, p_{x}, \phi_t, \phi_{1t}$ in terms of the prolonged system, we get more than 100 determining equations for the functions $X, T, U, \Phi, \Phi_1$ and $P$. Calculated by computer algebra, we finally get the desired result
\begin{eqnarray}\label{sol}
\nonumber&&X= \frac{c_1}{5}x+c_4,\, T= c_1t+c_2, \,U =-\frac{2c_1}{5}u+c_3\phi_1\phi,\, \Phi=\frac{1}{2}\phi(\frac{c_3}{2}p+c_5-\frac{c_1}{5}),\\&& \Phi_1 = \phi_1(-\frac{3c_1}{10}+\frac{c_3}{4}p+\frac{c_5}{2})+\frac{c_3}{4}\phi^3,\, P=\frac{c_3}{4}p^2+c_5p+c_6,
\end{eqnarray}
with arbitrary constants $c_1, c_2, c_3, c_4, c_5, c_6.$ When $c_3=2$ and $c_1=c_2=c_4=c_5=c_6=0$, the obtained symmetry degenerated into the special form in Eqs. \eqref{pointsy1} and \eqref{sigp}.
Consequently, the symmetries in \eqref{sigmasy} can be written explicitly as
\begin{eqnarray}\label{sigmauvf}
\nonumber\sigma_{u}&=&(\frac{c_1}{5}x+c_4)u_x+(c_1t+c_2)u_t+\frac{2c_1}{5}u-c_3\phi_1\phi,\\
\nonumber\sigma_{\phi}&=&(\frac{c_1}{5}x+c_4)\phi_x+(c_1t+c_2)\phi_t-\frac{1}{2}\phi(\frac{c_3}{2}p+c_5-\frac{c_1}{5}),\\
\nonumber\sigma_{\phi_1}&=&(\frac{c_1}{5}x+c_4)\phi_{1,x}+(c_1t+c_2)\phi_{1,t}-\phi_1(-\frac{3c_1}{10}+\frac{c_3}{4}p+\frac{c_5}{2})-\frac{c_3}{4}\phi^3,\\
\nonumber\sigma_{p}&=&(\frac{c_1}{5}x+c_4)p_x+(c_1t+c_2)p_t-\frac{c_3}{4}p^2-c_5p-c_6.
\end{eqnarray}
It can be seen from Eq. \eqref{sigmauvf} that the $c_3$-related symmetry is just the Lie point symmetry \eqref{pointsy1} and \eqref{sigp} obtained in the last section.

To give the group invariant solutions of the enlarged system, we have to solve the equations \eqref{sigmauvf} under symmetry constraints $\sigma_u=\sigma_{\phi}=\sigma_{\phi_1}=\sigma_p=0$, which is equivalent to solving the corresponding characteristic equation
\begin{equation}\label{chac}
\frac{dx}{X}=\frac{dt}{T}=\frac{du}{U}=\frac{d\phi}{\Phi}
=\frac{d\phi_1}{\Phi_1}=\frac{dp}{P}.
\end{equation}

Without loss of generality, we consider the nontrivial symmetry reductions ($c_3\neq0$) in the following two cases:

\noindent \textbf{Case 1}. $c_i\neq0$ (i=1,\,2,\,3,\,4,\,5, \,6).

Firstly, we make the transformations $c_3\rightarrow c_1c_3, c_5\rightarrow c_1c_5, c_6\rightarrow -c_1c_6$ and denote $\delta=\sqrt{c_6c_3+c_5^2} $ for computing simplicity. When $\delta\neq0$, solving equation \eqref{chac}, we have
\begin{eqnarray}\label{redusol01}
p&=&-\frac{2}{c_3}\big[c_5+\tanh(\frac{1}{2}\delta(\ln(c_1t+c_2)+c_1P))\delta\big],\\
\label{redusol03}\phi&=&(c_1t+c_2)^{(-\frac{1}{10})}\exp(-\frac{1}{10}c_1P)\Phi\sech(\frac{1}{2}\delta(\ln(c_1t+c_2)+c_1P))
,\\
\label{redusol05}\phi_1&=&-\frac{1}{\delta[\exp[\delta(\ln(c_1t+c_2)+c_1P)]+1]}(c_1t+c_2)^{(-\frac{3}{10})}
\exp(-\frac{3}{10}c_1P)
\nonumber\\&&[-\delta(\exp(\delta(\ln(c_1t+c_2)+c_1P))+1)\Phi_1+\Phi^3c_3]\sech(\frac{1}{2}\delta(\ln(c_1t+c_2)+c_1P))
,\\
\label{redusol06}u&=&\frac{1}{(c_1t+c_2)^{(\frac{2}{5})}\delta^2(\exp(\delta(\ln(c_1t+c_2)+c_1P))+1)^2}
\big\{2c_3^2\Phi^4\exp(-\frac{2c_1}{5}P)\nonumber\\&&-4c_3\Phi_1\delta[\exp(-\frac{2c_1}{5}P)+(c_1t+c_2)^{\delta}
\exp(\frac{c_1}{5}P(-2+5\delta))]\Phi\nonumber\\&&+\delta^2[2\exp(\delta(\ln(c_1t+c_2)+Pc_1))+1
+\exp(2\delta(\ln(c_1t+c_2)+Pc_1))]U\big\}.
\end{eqnarray}
Here, $P$, $\Phi,$ $\Phi_1$ and $U$ are all group invariant functions with group invariant variable $\xi=\frac{(c_1x+5c_4)}{(c_1t+c_2)^{\frac{1}{5}}c_1}$.

To get the symmetry reduction equations, we substitute Eqs. \eqref{redusol01}-\eqref{redusol05} into Eqs. \eqref{lax1xx}, \eqref{phi1} and \eqref{px} to find
\begin{equation}\label{redpphi}
P_{\xi}c_1\delta^2+\Phi^2\exp(-\frac{c_1}{5}P)c_3=0,
\end{equation}
\begin{equation}\label{redgphi_1}
c_3(1+5\delta)\Phi^3+10\Phi_{\xi}\delta^2\exp(\frac{c_1}{5}P)-10\Phi_1\delta^2=0,
\end{equation}
and
\begin{multline}\label{redhphi_1}
5\Phi^2c_3\Phi_1\delta\exp(-\frac{3}{10}c_1P)-10\Phi U\delta^2\exp(\frac{1}{10}c_1P)+10\exp(-\frac{1}{10}c_1P)\Phi_{1,\xi}\delta^2\\+3\Phi^2c_3\Phi_1
\exp(-\frac{3}{10}c_1P)=0,
\end{multline}
where $P$ satisfied the five-order ordinary differential equation   
\begin{multline}\label{redphi_1}
15c_1^4c\delta^4P_{\xi}^8-100c_1^2c\delta^2P_{\xi\xi\xi}P_{\xi}^5+(50c_1^2c\delta^2P_{\xi\xi}^2
+12c_1\xi)P_{\xi}^4+(-60+40cP_{\xi\xi\xi\xi\xi})P_{\xi}^3\\-(200cP_{\xi\xi\xi\xi}P_{\xi\xi}+100cP_{\xi\xi\xi}
^2)P_{\xi}^2+500cP_{\xi\xi\xi}P_{\xi\xi}^2P_{\xi}-225cP_{\xi\xi}^4=0.
\end{multline}

It is clearly that when $P$ is solved out from equation \eqref{redphi_1} new
group invariant solutions of the FOKdV equation \eqref{5kdv} would be immediately obtained through Eq. \eqref{redusol06}, where $\Phi,$ $\Phi_1$ and $U$ can be solved out from Eqs. \eqref{redpphi}, \eqref{redgphi_1} and \eqref{redhphi_1}. The entrance of $\tanh$ part in \eqref{redusol01} indicates that the interaction solution between nonlinear waves and solitons can thus be obtained. This type of interaction solutions are hard to be obtained by other approaches.

When $\delta=0$, repeating the similar steps as above, we get the explicit solutions for the FOKdV equation as
\begin{multline}
u=\frac{1}{4P_{\xi}^2(\ln(c_1t+c_2)+Pc_1)^2(c_1t+c_2)^{\frac{2}{5}}}\big[8P_{\xi}^4c_1^2-8c_1(\ln(c_1t+c_2)
+Pc_1)P_{\xi\xi}P_{\xi}^2\\+2(\ln(c_1t+c_2)+Pc_1)^2P_{\xi\xi\xi}P_{\xi}-(\ln(c_1t+c_2)+Pc_1)^2P_{\xi\xi}
^2\big],
\end{multline}
with $P$ satisfies the following equation
\begin{multline}
12P_{\xi}^4\xi c_1+(-60+40cP_{\xi\xi\xi\xi\xi})P_{\xi}^3+(-200cP_{\xi\xi\xi\xi}P_{\xi\xi}-100cP_{\xi\xi\xi}
^2)P_{\xi}^2+500cP_{\xi\xi\xi}P_{\xi\xi}^2P_{\xi}-225cP_{\xi\xi}^4=0.
\end{multline}

\noindent \textbf{Case 2}. $c_i\neq0$ (i=2,\,3,\,4,\,5,\,6) and $c_1=0$.

In this case, with out of generality, we make the transformations $c_6\rightarrow -c_6$ for computing simplicity. When $\delta\neq 0,$ following the similar procedure as in case1, we obtain the symmetry reduction solutions and corresponding symmetry reduction equations. Omitting these details, we give the final solution for the FOKdV equation
\begin{multline}
u=\frac{1}{4P_{z}^2c_2^2(\exp(\frac{\delta}{c_2}(t+P))+1)^2}\big[-\delta^2
(6\exp(\frac{\delta}{c_2}(t+P))-\exp(\frac{2\delta}{c_2}(t+P))-1)P_{z}^4\\-4c_2
\delta(-1+\exp(\frac{2\delta}{c_2}(t+P)))P_{zz}P_{z}^2+2c_2^2(\exp(\frac{2\delta}{c_2}(t+P))
+2\exp(\frac{\delta}{c_2}(t+P))+1)P_{zzz}P_{z}\\-c_2^2(\exp(\frac{2\delta}{c_2}(t+P))
+2\exp(\frac{\delta}{c_2}(t+P))+1)P_{zz}^2\big],
\end{multline}
where $z=\frac{c_2x-tc_4}{c_2}$ is the group invariant variable and the invariant function $P$ satisfies
\begin{multline}
3cP_{z}^8\delta^4-20\delta^2cc_2^2P_{zzz}P_z^5+(12c_2^3c_4+10\delta^2cc_2^2P_{zz}^2)P_z^4
+(8cP_{zzzzz}c_2^4-12c_2^4)P_z^3\\+(-20cP_{zzz}^2c_2^4-40cP_{zzzz}c_2^4P_{zz})P_z^2+100cP_zP_{zzz}
c_2^4P_{zz}^2-45cP_{zz}^4c_2^4=0.
\end{multline}

When $\delta=0,$ similar to the above case, we obtain exact solutions for the FOKdV equation, which reads
\begin{multline}
u=\frac{1}{4c_2^2P_z^2(\exp(\frac{\sqrt{2}c_5}{c_2}(t+P))+1)^2}\bigg[2c_5^2(-6\exp(\frac{\sqrt{2}c_5}{c_2}(t+P))
\\+\exp(\frac{2\sqrt{2}c_5}{c_2}(t+P))+1)P_z^4-4c_5\sqrt{2}c_2(\exp(\frac{\sqrt{2}c_5}{c_2}(t+P))-1)
(\exp(\frac{\sqrt{2}c_5}{c_2}(t+P))+1)P_{zz}P_z^2\\+2c_2^2(\exp(\frac{\sqrt{2}c_5}{c_2}(t+P))+1)^2P_{zzz}
P_z-c_2^2(\exp(\frac{\sqrt{2}c_5}{c_2}(t+P))+1)^2P_{zz}^2\bigg],
\end{multline}
with $P$ satisfying the following reduction equation
\begin{multline}
12cP_z^8c_5^4-40cP_z^5c_5^2P_{zzz}c_2^2+(20cc_5^2P_{zz}^2c_2^2+12c_2^3c_4)P_z^4+(8cP_{zzzzz}c_2^4
-12c_2^4)P_z^3\\+(-20cP_{zzz}^2c_2^4-40cP_{zzzz}c_2^4P_{zz})P_z^2+100cP_zP_{zzz}c_2^4P_{zz}^2
-45cP_{zz}^4c_2^4=0.
\end{multline}

\section{Conclusion and discussion}
In summary, the Lax pair related nonlocal symmetry has been obtained and used for deriving new exact solutions for the FOKdV equation. For that end, multiple new dependent variables were introduced to localize it in a new enlarged system. On this basis, new B\"{a}cklund transformation is obtained by using Lie's first theorem, from which new exact solutions can be derived from trivial seed solutions. Moreover, the general form of Lie point symmetry for the enlarged FOKdV system, which includes the localized Lax pair related nonlocal symmetry as a special case, is successfully obtained and the corresponding symmetry reduction solutions were explored. Two cases of symmetry reductions were obtained, from which new interaction solutions for the FOKdV equation could be obtained.

In this paper, we have shown that nonlocal symmetry can be used for constructing new exact solutions of the FOKdV equation, so it would be meaningful to get as much nonlocal symmetries as possible. Besides the method using Lax pair to derive nonlocal symmetries, there exist some other methods to obtain nonlocal symmetries, such as those obtained from B\"{a}cklund transformation\cite{97lou}, the bilinear forms and negative hierarchies, the nonlinearizations \cite{cao,cheng} and self-consistent sources \cite{zeng} etc. Using these various nonlocal symmetries to construct new interaction solutions for integrable systems deserves more deep investigation.

\begin{acknowledgments}
This work was supported by the National Natural Science Foundation of China under Grant Nos. 11347183, 11405110, 11275129 and 11305106, the Natural Science Foundation of Zhejiang Province of China Grant Nos. Y7080455 and LQ13A050001.
\end{acknowledgments}

\end{document}